


\overfullrule=0pt
\nopagenumbers
\magnification=1100
\voffset -1.2truecm  
\vsize=26truecm
\hsize=17.2truecm

\def\bls#1{\baselineskip=#1truept 
}

\def\title#1{\vbox to 2.5truecm{\parindent=0pt{#1}\vfil}  }
\def\authors#1{\vbox to 4.6truecm{\parindent=0pt{#1}\vfil}  }
\def\header#1{
$ $

\noindent {\bf #1} }
\def\subheader#1{
$ $

\noindent\underbar{\bf #1} }


\def \hb {\hfill\break}

\def\lessapprox{\hbox{$\lower1pt\hbox{$<$}\above-1pt\raise1pt\hbox{$\sim$}$}}
\def\greaterapprox{\hbox{$\lower1pt\hbox{$>$}\above-1pt\raise1pt\hbox{$\sim$}$}}

\def\hb{\hfil\break}

\bls{16}
$ $
\vskip 1.8truecm

\title{
\underbar{
EXTRACTING INFINITE SYSTEM PROPERTIES FROM FINITE SIZE CLUSTERS:
}\hfil\break\underbar{
``PHASE RANDOMIZATION/BOUNDARY CONDITION AVERAGING''
}}
\authors{
{J.~Tinka GAMMEL$^{(a)}$}, D.K.~CAMPBELL$^{(b)}$,
and E.Y.~LOH, Jr.$^{(c)}$\hb
$^{(a)}$ Code 573, Materials Research Branch,
NCCOSC RDT\&E Division (NRaD),
San Diego, CA 92152-5000, USA \hb
$^{(b)}$Dept.~of Physics, University of Illinois at Urbana-Champaign,
1110 W.~Green St., Urbana, IL 61801, USA  \hb
$^{(b)}$Thinking Machines Corporation, 1010 El Camino Real, Suite 310,
Menlo Park, CA 94025, USA
}
\header{ABSTRACT}

When electron-electron correlations are important, it is often necessary to
use ``exact'' numerical methods, such as Lancz\"os diagonalization, to study
the
full many-body Hamiltonian. Unfortunately, such exact diagonalization
methods are restricted to small system sizes.
We show that if the Hubbard $U$ term is replaced by a ``periodic Hubbard" term,
the full many body Hamiltonian may be exactly solved, even for very large
systems, though for low fillings. However, for half-filled systems
and large $U$ this approach is not only no longer exact, it no longer
improves extrapolation to larger systems.
We discuss how generalized ``randomized variable averaging'' (RVA)
or  ``phase randomization'' schemes can be reliably employed to improve
extrapolation to large system sizes in this regime.
This general approach can be combined
with any many-body method and is thus of broad interest and applicability.

\header{INTRODUCTION}

Solutions of the Peierls-Hubbard
Hamiltonian (PHH) [1] for small clusters
are strongly dependent on the boundary condition (BC); $i.e.$, for the
case of polyacetylene, whether
the carbon atoms are viewed as being on a real ring, such as benzene,
a real chain, such as (1,3,5)-hexatriene, or a more exotic geometry,
such as anti-periodic BCs.
We have found that
[2,3] to reduce finite size corrections
to the calculated  optical gaps and spectra, thus
improving the extrapolation to the infinite case, it is effective
to employ novel ``averaging"
techniques to ``randomize" the many-body levels which influence
those correlation functions of interest.
Briefly, by randomization, we mean simply that the total
energy of the system is to be viewed as a weighted average of the
energies derived separately for each of several different
``random" values (RV) of a particular parameter,
which may be the BC,
a hopping integral, an on-site energy, or the Hubbard $U$:
$ E=\sum_{\{ RV\} } x_{RV} E[RV] $,
where $x_{RV}$ is a normalized weighting factor with
$\sum_{\{ RV\} } x_{RV} =1$.
Similarly, any correlation function can be
viewed as the average of of its values from the separate RVs.
The number of RV, $N_{RV}$, may be varied,
and the choice of the RV and $x_{RV}$
may be truly random, or may be a pre-specified set of values,
which may in turn depend on the quantity being studied.
The case of {\it phase} randomization and its relation
to Bloch's theorem are described below.
More generally, this randomized-variable-averaging (RVA) technique
may be viewed as a procedure to mimic disorder.

We focus in this manuscript
on the calculation via exact Lancz\"os diagonalization
of the self-consistent uniform dimerization
and one-photon optical absorption of the ground state
within the 1-D, 1/2-filled, one-band PHH.
However, RVA is equally applicable to the calculation of, $e.g.$, the
phonon modes of the ground state,
as well as the geometry and self-consistent absorptions of
doped or higher lying ($e.g.$, triplet) states [3],
luminescence spectra, structure factors, $etc.$,
in single- or multi-band models in 1-, 2-, and 3-D,
via Lancz\"os diagonalization, as well as
other numerical procedures on small lattices, such as Monte Carlo.


\header{PHASE BOUNDARY CONDITION AVERAGING AND BLOCH'S THEOREM}

Bloch's theorem tells us that the
single particle wavefunctions of a system of size $M$$\cdot$$N$ with
periodicity $N$  and periodic boundary conditions (PBCs)
are exactly the single particle wavefunctions of the ensemble
of systems of size $N$ for each of the phase BCs
$\psi (N+1)=e^{i\phi}\psi(1)$, $\phi=2\pi\ell/M$, $\ell=1,...,M$.
Restated, the (exact) properties of a larger
(single particle) system may be found by forming a (particular) average over
smaller systems with different BCs.
This ``boundary condition averaging" (BCA)
or ``phase randomization"
is a special case of our general RVA procedure.
Blochs theorem may be generalized to many particle wavefunctions
and is applicable here if the form of the Hamiltonian
is modified slightly.
The one-dimensional PHH is
$$
H~=\sum_{\ell,\sigma} (-t_0+\alpha\delta_\ell)
(c^\dagger_{\ell~\sigma} c_{\ell+1~\sigma}
+c^\dagger_{\ell+1~\sigma} c_{\ell~\sigma})~
+~ U \sum_\ell n_{\ell\uparrow} n_{\ell\downarrow}~
+~ {1\over 2} K\sum_\ell \delta_\ell^2
\eqno{(1)}
$$
If we replace the Hubbard term,
$ {U} \sum_{\ell=1}^{N \cdot M}
 n_{\ell,\uparrow} n_{\ell,\downarrow}, $
by its periodic analog,
$$
H_{el-el}~\rightarrow~{U \over M} \sum_{\mu,\nu=1}^M ~\sum_{\ell=1}^N
 n_{\ell+\mu\cdot N,\uparrow} n_{\ell+\nu\cdot N,\downarrow} ~,
\eqno{(2)}
$$
then, if the original problem had periodicity $N$,
$\langle n_{\ell,\sigma}\rangle = \langle n_{\ell+\nu\cdot N,\sigma}\rangle$,
we have, in a
mean-field sense, the ``same'' Hamiltonian. With this new ``periodic-Hubbard''
Hamiltonian (periodic PHH)
the Bloch analysis of a given large (periodic) system
($N\cdot M$ sites, $N_e \cdot M$ electrons)
is accomplished by considering a small system
($N$ sites, $N_e$ electrons)
with several different BCs. The method scales linearly
with $M$, allowing one to handle reasonably large systems, and sets a sound
theoretical basis for the empirical observation [2] that
RVA can be used to smooth optical absorption
spectra obtained via exact finite-size diagonalization.

Using the periodic-Hubbard term, Eq.~(2), assuming the lattice
distortion has the same periodicity,
$\delta(\ell+N)=\delta(\ell)$, and using
$\sum_{\mu=1}^M \sum_{\ell=1}^N \sum_{\sigma=\uparrow,\downarrow}
n_{\ell+\mu\cdot N,\sigma} \equiv N_e \cdot M$,
one can show:
$$
H_{per}~=~{1\over M}~\sum_{\mu,\nu=1}^M ~\sum_{i,j,k,l=1}^N
             ~\sum_{s_1,s_2=\uparrow,\downarrow}~
 c_{i+\mu\cdot N,s_1}^\dagger  c_{j+\mu\cdot N,s_1}  ~
\times  ~ H(i,j,k,l;s_1,s_2)~
 c_{k+\nu\cdot N,s_2}^\dagger c_{l+\nu\cdot N,s_2}
\eqno{(3)}
$$
where
$$\eqalign{
H(i,j,k,l;s_1,s_2)=&H(i+N,j+N,k,l;s_1,s_2)  \cr
=&H(i,j,k,l;s_2,s_1)  ~=~H(k,l,i,j;s_1,s_2)  ~=~H(j,i,k,l;s_1,s_2)   
}$$
and
$$\eqalign{
H(i,j,k,l;s_1,s_2)~~=&~~{1\over 2} \biggm[{\sum_{n=1}^N (-t_0+\alpha\delta_n)
(\partial_{i,n}\partial_{j,n+1}+\partial_{i,n+1}\partial_{j,n})
{\partial_{k,l}\over{N_e}} ~~+~~(i,j \rightleftharpoons k,l) }\biggm]  \cr
&~~+~~  {1\over 2} K \sum_{n=1}^N \delta_n^2
                ~{\partial_{i,j}\over{N_e}}{\partial_{k,l}\over{N_e}}~~
+~~ U \partial_{s_1,\bar s_2}\partial_{i,j}\partial_{k,l} ~~.
}$$
Note that $H(i,j,k,l;s_1,s_2)$ is independent of $\mu$, $\nu$, and $M$.
The Bloch analysis on the many-body eigenfunctions using the
symmetries of $H(i,j,k,l;s_1,s_2)$ leds to a $\Psi$ of the form:
$$\eqalign{
\Psi_{k_1,...,k_p}~=\sum_{\sigma_1,...,\sigma_p=\uparrow,\downarrow}
&~\sum_{n_1,...,n_p=1}^N ~\psi_{k_1,...,k_p}(n_1,...,n_p)  \cr
\times & ~{ \sum_{\mu_1=1}^M c_{n_1+\mu_1\cdot N,\sigma_1}^\dagger
                  e^{ik_1\mu_1} } ~\times ~\cdot\cdot\cdot~
\times  ~{ \sum_{\mu_p=1}^M c_{n_p+\mu_p\cdot N,\sigma_p}^\dagger
                  e^{ik_p\mu_p} }
}\eqno{(4)}$$
where
$\psi_{k_1,...,k_\ell,...,k_p}(n_1,...,n_\ell,...,n_p)
=\psi_{k_1,...,k_\ell,...,k_p}(n_1,...,n_\ell+\mu_\ell\cdot N,...,n_p)~$,
$k_\ell=(2\pi j/M)$, $~j\epsilon\{ 0,...,M-1\}$,
and $p\equiv N_e\cdot M$.  We find for $i,j\epsilon\{1,...,N\}$,
$$\eqalign{
\sum_{\nu_1=1}^M  c_{i+\nu\cdot N,s}^\dagger
c_{j+(\nu+\nu_0)\cdot N,s} &\Psi_{k_1,...,k_p}~
=\sum_{\sigma_1,...,\sigma_p=\uparrow,\downarrow}~
\sum_{n_1,...,n_p=1}^N ~\sum_{\ell=1}^N ~\psi_{k_1,...,k_p}(n_1,...,n_p)~
\partial_{n_\ell,j}\partial_{\sigma_\ell,s}~e^{ik_\ell\nu_0}  \cr
\times & { \sum_{\mu_1=1}^M c_{n_1+\mu_1\cdot N,\sigma_1}^\dagger
                  e^{ik_1\mu_1} }
\times \cdot\cdot\cdot  \times
{ \sum_{\mu_\ell=1}^M c_{i+\mu_\ell\cdot N,\sigma_\ell}^\dagger
                  e^{ik_\ell\mu_\ell} }
\times \cdot\cdot\cdot  \times
{ \sum_{\mu_p=1}^M c_{n_p+\mu_p\cdot N,\sigma_p}^\dagger
                  e^{ik_p\mu_p} }
}\eqno{(5)}$$
($\nu_0$ comes into play for operators like
$c_{N,s}^\dagger c_{N+1,s}$). From Eq.~(5) we see
$H_{per}$ is diagonal in the $k$'s,
$$
H_{per}~\Psi_{k_1,...,k_p}^\beta~=~E_\beta~\Psi_{k_1,...,k_p}^\beta~~,
\eqno{(6)}
$$
as are
$\rho=\sum_{\ell,\sigma} n_{\ell,\sigma}$
and $J=i\sum_{\ell,\sigma}(-t_0  + \alpha \delta_{\ell})
     ( c_{\ell,\sigma}^\dagger c_{\ell+1,\sigma}
          -c_{\ell+1,\sigma}^\dagger c_{\ell,\sigma})$,
and the eigenfunctions of $H_{per}$ can be written in the form Eq.~(4).

The symmetry of $\Psi$ implies that if $\sigma_{\ell_1}=\sigma_{\ell_2}$
and $k_{\ell_1}=k_{\ell_2}$ then $n_{\ell_1}\ne n_{\ell_2}$. Thus there
are at most $2N$ of the $k_\ell$ the same. We postulate
that the ground state lies in the manifold
with each of the $M$ distinct values for $k_\ell$ occurring $N_e$ times
and (for the half-filled band) with equal numbers of up and down spins
(in general, the largest manifold).
This can be checked, and we stress that the analysis
up to this point is exact for the periodic PHH. In this manifold
we can write $\Psi$ as:
$$
\Psi^\beta ~=~\sum_{\alpha_1,...,\alpha_M=1}^{2N}~
\phi^\beta(\alpha_1,...,\alpha_M)~
\Psi_{q_1,\alpha_1}^{N_e}~\cdot\cdot\cdot~
\Psi_{q_M,\alpha_M}^{N_e} ~,
\eqno{(7)}
$$
where $q_\ell \equiv (2\pi\ell/M)$ and
$$\eqalign{
\Psi_{q,\alpha}^{N_e}~=~\sum_{\sigma_1,...,\sigma_{N_\uparrow}=\uparrow}
&~\sum_{\sigma_{N_\uparrow+1},...,\sigma_{N_e}=\downarrow}~
\sum_{n_1,...,n_{N_e}=1}^N ~\psi_{q,\alpha}^{N_e}(n_1,...,n_{N_e})  \cr
\times & ~{ \sum_{\mu_1=1}^M c_{n_1+\mu_1\cdot N,\sigma_1}^\dagger
                  e^{iq_1\mu_1} } ~\times ~~\cdot\cdot\cdot~~
\times  ~{ \sum_{\mu_{N_e}=1}^M
                c_{n_{N_e}+\mu_{N_e}\cdot N,\sigma_{N_e}}^\dagger
                  e^{iq_{N_e}\mu_{N_e}} }
}$$
where $N_\uparrow={[[} {{(N_e+1)}\over 2} {]]}$,
$N_\downarrow={[[}{N_e\over 2}{]]}$ ~(${[[}~{]]}$
denotes integer value of). ~Note
$$
c_{1+N,\sigma}^\dagger c_{N,\sigma}~ \Psi_{q,\alpha}^{N_e} ~=~
e^{iq} ~c_{1,\sigma}^\dagger c_{N,\sigma}~ \Psi_{q,\alpha}^{N_e} ~,
\eqno{(8)}
$$
and so solving the $N_e$ electron problem
on the full $N\cdot M$ sites with PBCs for fixed $q$ is {\it exactly}
equivalent
to solving
$$
H_{per}~\Psi_{q,\alpha}^{N_e}~=~E_{q,\alpha}^{N_e}~\Psi_{q,\alpha}^{N_e}
\eqno{(9)}
$$
on $N$ sites with the $q$-dependent BC defined by Eq.~(8).
Thus, to find the exact $N_e$ electron eigenfunctions $\Psi_{q,\alpha}^{N_e}$
on $N\cdot M$ sites,
we only need to solve Eq.~(9) with $N$ sites and $N_e$ electrons
for each of the $M$ BCs,
rather than the full $N\cdot M$ site problem.

To obtain the exact eigenfunctions of the full $N_e$$\cdot$$M$ electron
problem,
we need to solve for $\phi^\beta(\alpha_1,...,\alpha_M)$. This is where
an approximation must be made to be able to solve the problem
numerically, as we assume $N$ is already as large as computationally
feasible. The full eigenvalue problem yields
$$\eqalign{
\biggm[{H_{per}~,~\Psi_{q,\alpha}^{N_e}}\biggm]~=~E_{q,\alpha}^{N_e}~
+~{U\over M} ~\sum_{\ell=1}^N ~\biggm\{ ~~
&\biggm[{\sum_{\mu=1}^M n_{\ell+\mu\cdot N,\uparrow}~,~
     \Psi_{q,\alpha}^{N_e}}\biggm] \sum_{\nu=1}^M
n_{\ell+\nu\cdot N,\downarrow}   \cr
+~&\biggm[{\sum_{\nu=1}^M n_{\ell+\nu\cdot N,\downarrow}~,~
     \Psi_{q,\alpha}^{N_e}}\biggm] \sum_{\mu=1}^M n_{\ell+\mu\cdot
N,\uparrow}~~
\biggm\} 
}\eqno{(10)}$$
To solve Eq.~(6) we must use some approximation for Eq.~(10),
such as perturbation theory or mean-field. Thus BCA
means we treat electron-electron
correlations within the $N_e$ electron manifold exactly,
and electron-electron correlations between $N_e$ electron manifolds
approximately. The BCA results reported
here are zero-order perturbation theory: we have assumed
$\phi^\beta (\alpha_1,...,\alpha_M)$ to be a product of $\delta$-functions;
$i.e.$, for the ground state we use:
$$
\Psi^0~\simeq~ ~\Psi_{q_1,0}^{N_e}~\cdot\cdot\cdot~
              ~\Psi_{q_M,0}^{N_e}  ~;~~~~~
E_0~\simeq~\sum_{\ell=1}^M~E_{q_\ell,0}^{N_e}~.
\eqno{(11)}
$$
Note that if we had been interested in systems at low filling
($N_e$ electrons),
rather than near half filling ($N_e\cdot M$ electrons),
there would be no need to make any approximations.

\subheader{Comparison of the periodic and standard Peierls-Hubbard
Models at half-filling}

At $U$=0 Bloch's theorem for the many particle wavefunctions
as formulated above is exact at all fillings.
For the half-filled PHH, we have just shown
that phase BCA involves two approximations:
first, replacing the Hubbard term by Eq.~(2), and, second,
ignoring correlations between the $N_e$ particle wavefunction
leading to Eq.~(11). We now test these approximations.
In Fig.~1 we show the minimum energy dimerization of an 8-site system
as a function of $U$ for ($i$) phase averaged solution and ($ii$)
the exact solution of the half-filled band with the periodic
Hubbard term and periodicity 2,4, and 8 (periodicity 8 is the
the usual Hubbard).  The approximate periodic
(BCA) solution lies between the exact
periodic solution and the exact pure-Hubbard solution.
Since the original problem was for
$U\cdot M \sum_{\ell=1}^{N\cdot M} n_{\ell\uparrow}
n_{\ell\downarrow}$ on $N \cdot M$
sites, one might argue that the approximate phase-averaged result for
$U \sum_{\ell=1}^{N} n_{\ell\uparrow} n_{\ell\downarrow}$ on $N$
sites may be closer to the desired answer than the exact periodic-Hubbard
$N\cdot M$ site result.

\topinsert \vglue 3.2truein \bls{12} \noindent
Fig.~1 (left).
Comparison of the approximate and exact periodic-Hubbard
minimum energy dimerization as a function
of $U$ for
$t_0$=2.5 eV, $\alpha$=4.1 eV/\AA, and $K$=21 eV/\AA$^2$.
\vskip 12truept \noindent
Fig.~2 (right).
Dimerization $vs.$ $U$ for various $N$: phase BCA.
Parameters as Fig.~1.
\endinsert \bls{16}

In Fig.~2 we show the dimerization amplitude obtained by phase BCA.
We see that the extrapolated infinite behavior
at small to intermediate $U$ is well approximated
after phase averaging for $N$=10 even without extrapolation.
The agreement is best for small $U$, while no change
from the pure periodic behavior is found for large $U$.
That this particular technique
is expected to have no effect at large $U$ can be seen by examining
the effective spin-Peierls Hamiltonian [4],
which is asymptotically independent of the phase of the BC.
However, the phase-averaged analysis does lead to the interpretation
of the finite-size
result as the result for the larger system with an effective periodic-Hubbard
interaction. Thus, if $U$ is scaled by the
system size, as is done in Fig.~2,
the infinite behavior should be more easily extrapolated.
It appears that this is at least approximately true.

To show how phase BCA affects the optical spectrum.
in Fig.~3a
we show the (Lorentzian broadened) spectrum at $U/t_0$=0.4
for $N$=8 and 5 phase BCs (enough to yield a smooth spectrum at $U$=0).
It is clear how the different BCs ``fill in" the spectrum.
Fig.~3b shows the spectrum at $U/t_0$=1.6.
Here, despite the several BCs used,
the spectrum remains sparse, due to the loss of effectiveness of the
phase BCA scheme as discussed above.
If we were to use a broad enough Lorentzian to smooth this spectrum
(the same width as in Fig.~3a was used), the optical gap edge
would be lost.

\topinsert \vglue 3.2truein \bls{12} \noindent
Fig.~3.
The Lorentzian broadened
optical absorption spectra
calculated using the
phase BCA method for (a) $U$=1 eV
and (b) $U$=4 eV.
Note the ``sparseness''
of the spectrum in the latter case indicating the failure of the
phase BCA approach.
For comparison, we also show
the spectrum at $U$=10 eV using (c) amplitude BCA and (d)
an RVA procedure where $U$ is varied on one site.
Note these RVA spectra are dense and show the same features.
Parameters as Fig.~1 but $\delta$=0.14 \AA.
\endinsert \bls{16}

\header{AMPLITUDE BOUNDARY CONDITION AVERAGING}

The modification of the phase of the BC discussed above can
also be viewed as passing a random magnetic
flux through an ensemble of closed ring and studying average
properties. In this sense we have ``randomized'' the locations
of the momentum space states.
We have already indicated that RVA can help us
transcend the limitations of this ``bond phase/magnetic flux" approach.
We can ``randomize'' electronic properties by, $e.g.$, changing a local
hopping or an on-site energy or a Coulomb repulsion somewhere on the chain.
One could also introduce an additional field and vary it about zero.
Although there has been some limited earlier work on using
modified BCs in the context of the Hubbard
model [5], there is as yet no provably accurate prescription for
for arbitrary $U$ and $V$. However, certain intuitive rules must
guide us. First, whatever change is made to randomize must, of course, do so
effectively. Second, the change in the system must be negligible as
the lattice size is increased to infinity; for example,
if only one bond or site is varied from calculation to calculation, then
the effect of such a change is immaterial in the thermodynamic limit.
Finally, the behavior for small lattice sizes must be illustrative of
the infinite-size limit. Put another way,
one must still be able to make a reasonable extrapolation
to the infinite chain.

We have found [2] an ``amplitude BCA'' (``scaled-hopping'')
technique to be effective and produce results in good
agreement with expectations based on both strong- and weak-coupling
analytic arguments. For this method,
we randomize by varying the {\it magnitude} rather than the
phase of the ``boundary'' hopping
-- $i.e.$, between sites 1 and $N$ -- typically
from $-t_0$ to $+t_0$ in ten to twenty equal steps.
The individual spectra are then added together
with weights $x_i$ chosen to minimize the total length of the
final curve, though giving each spectra equal weight yields
virtually identical results.
This clearly incorporates the special case of using
only periodic ($x$=1) or antiperiodic ($x$=$-$1)
rings, the JT/nJT difference being important for weak coupling.
It also incorporates the case of the open chain ($x$=0), which
has the ``best'' single BC size dependence (though
with attendant ``end effect'' problems).
Finally, it works in the strong-coupling limit.

In Fig.~3c, we show the spectra produced by the amplitude BCA
technique for for a larger $U$ than that in Fig.~3b where
phase BCA failed. In Ref.~[2] we showed that
not only are the gross features of the spectra obtained by amplitude
BCA in agreement with
strong-coupling calculations [6], but
in addition they show substantially more interesting detail,
such as the  ``decoupled-dimer" peak
located at $\sim U/2+\sqrt{(U/2)^2+(2t_0)^2}$
in systems with strong electron-phonon {\it and} electron-electron couplings.
To emphasize the generality, in
Fig.~3d we show the spectrum obtained by varying $U$
on the first site between 0 and 20 eV.
This large $U$ variation produces states in the optical gap,
but otherwise the spectrum is unchanged from Fig.~3c.
We feel that these results, and those in Ref.s [2,3], demonstrate that RVA
is an effective means for reducing (and in some cases
practically eliminating) finite size dependence, yielding results that
can be confidently extrapolated to the infinite size limit, and which are
in good agreement with known analytic results.

\underbar{\it Acknowledgements}.
We wish to thank
J.~Bronzan, J.~Gubernatis, S.~Mazumdar, and D.~Scalapino, among others,
for many useful discussions.
JTG was supported by a
NRC-NRaD Research Associateship through a grant from
ONR. 
Computational support was provided by
LANL and TMC.

\header{REFERENCES}
\parskip=0pt

\item{1}
For a survey of the PHH applied to conducting polymers see
D.~Baeriswyl, D.K.~Campbell, and S.~Mazumdar,
in H.~Kiess (ed.), {\it Conducting Polymers}, (Springer, 1991).
\item{2}
E.Y.~Loh,~Jr., and D.K.~Campbell,
{\it Synth.~Metals}, {\bf 27} (1988) A499;
D.K.~Campbell, J.T.~Gammel, and E.Y.~Loh,~Jr.,
{\it Int.~J.~Mod.~Phys.~B}, {\bf 3} (1989) 2131;
E.Y.~Loh,~Jr., D.K.~Campbell, and J.T.~Gammel,
in D.~Baeriswyl and D.K.~Campbell (ed.),
{\it Proc.~of the NATO ARW on Interacting Electrons in
Reduced Dimensions, Torino, Italy, Oct.~3-7, 1988},
(Plenum, 1990).
\item{3}
D.K.~Campbell, J.T.~Gammel, H.Q.~Lin, and E.Y.~Loh,~Jr,
{\it Synth.~Metals}, {\bf 49}, (1992), in press; 
D.K.~Campbell, J.T.~Gammel, and H.Q.~Lin, this conference.
\item{4}
J.E.~Hirsch,
{\it Phys.~Rev.~Lett.}, {\bf 51} (1983) 296.
\item{5}
R.~Julien and R.M.~Martin, {\it Phys.~Rev.~B}, {\bf 26} (1982) 6173;
A.M.~Oles\', G.~Tr\'eglia, D.~Spanjaard, and R.~Julien,
{\it Phys.~Rev.~B}, {\bf 32} (1985) 2167.
\item{6}
S.K.~Lyo and J.-P.~Gallinar, {\it J.~Phys.~C}, {\bf 10} (1977) 1693;
S.K.~ Lyo, {\it Phys.~Rev.~B}, {\bf 18} (1978) 1854;
J.-P.~Gallinar, {\it J.~Phys.~C}, {\bf 12} (1979) L335.

\end